%% file: eprint.tex
\newcommand*{\ATLASLATEXPATH}{atlaslatex/}
\documentclass[UKenglish,10pt]{article}
\usepackage{float}
\usepackage{comment}
\usepackage{enumitem}
\usepackage{makecell}
\usepackage{amsmath,amssymb} %
\usepackage{multirow}
\usepackage{caption}
\usepackage[utf8]{inputenc} 
\usepackage{\ATLASLATEXPATH atlasbiblatex}

\usepackage{\ATLASLATEXPATH atlascontribute}

\usepackage[BSM,process,xref]{\ATLASLATEXPATH atlasphysics}

\usepackage{pdflscape}
\usepackage{adjustbox}

\graphicspath{{logos/}{figures/}}
\usepackage{paper-defs}
\def\Title#1{\begin{center} {\Large #1 } \end{center}}
\def\Author#1{\begin{center}{ \sc #1} \end{center}}
\def\Address#1{\begin{center}{ \it #1} \end{center}}

\newcommand\pubblock{\rightline{\begin{tabular}{l} Proceedings of the Fifth Annual LHCP\\ \pubnumber\\
         \pubdate  \end{tabular}}}

\newenvironment{Abstract}{\begin{quotation} \begin{center} 
             \large ABSTRACT \end{center}\bigskip 
      \begin{center}\begin{large}}{\end{large}\end{center} \end{quotation}}

\newenvironment{Presented}{\begin{quotation} \begin{center} 
             PRESENTED AT\end{center}\bigskip 
      \begin{center}\begin{large}}{\end{large}\end{center} \end{quotation}}

\input econfmacros.tex

\newcommand\pubnumber{ ATL-PHYS-PROC-2017-117 }

\newcommand\pubdate{\today}

\def\affiliation{
On behalf of the ATLAS Collaboration, \\
Department of Physics and Astronomy \\
University of Pennsylvania, Philadelphia, PA, USA}

\usepackage{lineno}

\begin{document}

\large
\begin{titlepage}
\pubblock

\vfill
\Title{  Search for supersymmetric partners of third generation quarks \\ in leptonic channels with the ATLAS detector }
\vfill

\Author{ Keisuke Yoshihara  }
\Address{\affiliation}
\vfill
\begin{Abstract}
Two of the most important parameters in supersymmetry are the masses of the stop and sbottom, the supersymmetric partners of the third generation quarks. A stop mass lighter than 1 TeV is favored theoretically; however, ``conventional'' searches based on the simplified models have not produced experimental evidence for a light stop. It is possible that the light stop evades our searches due to a compressed sparticle mass spectrum. Therefore, the searches are extended to cover a broader range of signal scenarios with different mass splittings between the stop, neutralino(s), and chargino(s). The searches are then interpreted in the context of both simplified models and pMSSM models. Searches are also performed for various $R$-parity violating stop (sbottom) models. Recent ATLAS results from searches for direct stop (sbottom) pair production are presented in final states with jets, missing transverse-momentum, and leptons. The analyses are based on 36 fb$^{-1}$ of $\sqrt{s}$=13 TeV proton-proton collision data recorded with ATLAS detector at the LHC in 2015 and 2016.
\end{Abstract}
\vfill

\begin{Presented}
The Fifth Annual Conference\\
 on Large Hadron Collider Physics \\
Shanghai Jiao Tong University, Shanghai, China\\ 
May 15-20, 2017
\end{Presented}
\vfill
\end{titlepage}
\def\thefootnote{\fnsymbol{footnote}}
\setcounter{footnote}{0}
%

\normalsize 


\section{Introduction}

The hierarchy problem~\cite{Weinberg:1975gm,Gildener:1976ai,Weinberg:1979bn,Susskind:1978ms} 
has gained additional attention with the observation of the Standard Model (SM) Higgs boson at the Large Hadron Collider (LHC)~\cite{LHC:2008}. 
Supersymmetry (SUSY)~\cite{Miyazawa:1966,Ramond:1971gb,Golfand:1971iw,Neveu:1971rx,Neveu:1971iv,Gervais:1971ji,Volkov:1973ix,Wess:1973kz,Wess:1974tw}, which extends the SM by introducing supersymmetric partners for every SM degree of freedom, can provide an elegant solution to the hierarchy problem. The dominant divergent contribution to the Higgs boson mass due to loop diagrams involving top-quarks can be largely cancelled by introducing the stop because of its large Yukawa coupling.
\vspace{0.20cm}

\hspace{-0.60cm}
The masses of the third-generation squarks, and in particular the mass of the stop, can be significantly lower than those of the other generations, with masses well within the reach of LHC. This is possible with large stop mixing (achievable due to the large top-quark Yukawa coupling) and because of the strong effect of the renormalisation group equations for the third-generation squarks. If the light stop is mostly composed of the left-handed state, a sbottom (the superpartner of the bottom quark) can also be light, as the masses of the two states are controlled by a common mass parameter at tree-level.
\vspace{0.20cm}

\hspace{-0.60cm}
General models of SUSY need not conserve baryon number (B) and lepton number (L), resulting in a proton lifetime shorter than current experimental limits. This is commonly resolved by introducing a multiplicative quantum number called $R$-parity, which is $1$ and $-1$ for all SM and SUSY particles, respectively. A generic $R$-parity-conserving minimal supersymmetric extension of the SM (MSSM) predicts pair production of SUSY particles and the existence of a stable lightest supersymmetric particle (LSP). 
Proton decay can be also avoided by allowing for either L or B violation. ATLAS conducts extensive SUSY searches assuming both $R$-parity conservation and violation.
\vspace{0.20cm}

\hspace{-0.60cm}
This article presents various searches (stop 1-lepton, stop 2-lepton, stop B-L, and RPV 1-lepton) based on newly available CONF notes and papers~\cite{ATL-CONF-2017-034,ATL-CONF-2017-036,ATL-CONF-2017-037,RPV1L} for direct \tone\ pair production in final states with isolated charged lepton(s) and jets including $b$-tagged jets. In $R$-parity conserving scenarios, the two weakly-interacting LSPs will escape detection, which can lead to significant missing transverse-momentum $\met$. On the other hand, $\met$ is expected to be small in the $R$-parity violating scenario, as the unstable LSPs decay into SM particles. 

\section{ATLAS detector}

The ATLAS detector~\cite{PERF-2007-01} is a multipurpose particle physics detector with nearly $4\pi$ coverage in solid angle around the collision point. It consists of an inner tracking detector (ID), surrounded by a superconducting solenoid providing a 2 T axial magnetic field, a system of calorimeters, and a muon spectrometer (MS) incorporating three large superconducting toroid magnets.
\vspace{0.20cm}

\hspace{-0.60cm}
The ID provides charged-particle tracking in the range $|\eta| < 2.5$. High-granularity electromagnetic and hadronic calorimeters cover the region $|\eta| < 4.9$. The central hadronic calorimeter is a sampling calorimeter with scintillator tiles as the active medium and steel absorbers. All of the electromagnetic calorimeters, as well as the endcap and forward hadronic calorimeters, are sampling calorimeters with liquid argon as the active medium and lead, copper, or tungsten absorbers. The MS consists of three layers of high-precision tracking chambers with coverage up to $|\eta|=2.7$ and dedicated chambers for triggering in the region $|\eta|<2.4$. Events are selected by a two-level trigger system: the first level is a hardware-based system and the second is a software-based system. 
\vspace{0.20cm}

\hspace{-0.60cm}
Analyses presented here are based on a dataset corresponding to a total integrated luminosity of 36.1 fb$^{-1}$ collected in 2015 and 2016 at a collision energy of $\sqrt{s} = 13$\,\TeV. The data contain an average number of simultaneous $pp$ interactions per bunch crossing, or ``pileup'', of approximately 23.7 across the two years. The events are primarily collected with \met\ or lepton triggers. The \met\ trigger is fully efficient for events where the offline-reconstructed $\met > 200$\,$\GeV$. This is the minimum \met\ required in all signal and control regions relying on the \met\ triggers. 

\section{$R$-parity conserved scenarios}

\subsection{Signal models}
The general analysis strategy is to probe a broad range of the possible realisations of SUSY scenarios, taking the approach of defining dedicated search regions to target specific but representative SUSY models. The phenomenology of each model is largely driven by the composition of its lightest supersymmetric particles, which are considered to be some combination of the electroweakinos: the supersymmetric partners of the SM gauge bosons and Higgs boson. In practice, this means that the most important parameters of the SUSY models considered are the masses of the electroweakinos and of the colour-charged third generation sparticles.
\vspace{0.20cm}

\hspace{-0.60cm}
Searches are conducted targeting signals described either by simplified models or the phenomenological MSSM (pMSSM) models. In simplified models, the masses of all sparticles are set to high values (``decoupled'') except for the few sparticles involved in the decay chain of interest. In pMSSM models, each of the 19 free pMSSM parameters are set to some fixed, physically-motivated values, except for two mass parameters which are scanned. The set of models used are chosen to give broad coverage of the possible phenomenology of stop decays that can be realised in the MSSM, in order to provide a general statement on the sensitivity of the search for direct stop production. 
\vspace{0.20cm}

\begin{figure}[htbp]
\begin{center}
\includegraphics[width=0.75\textwidth]{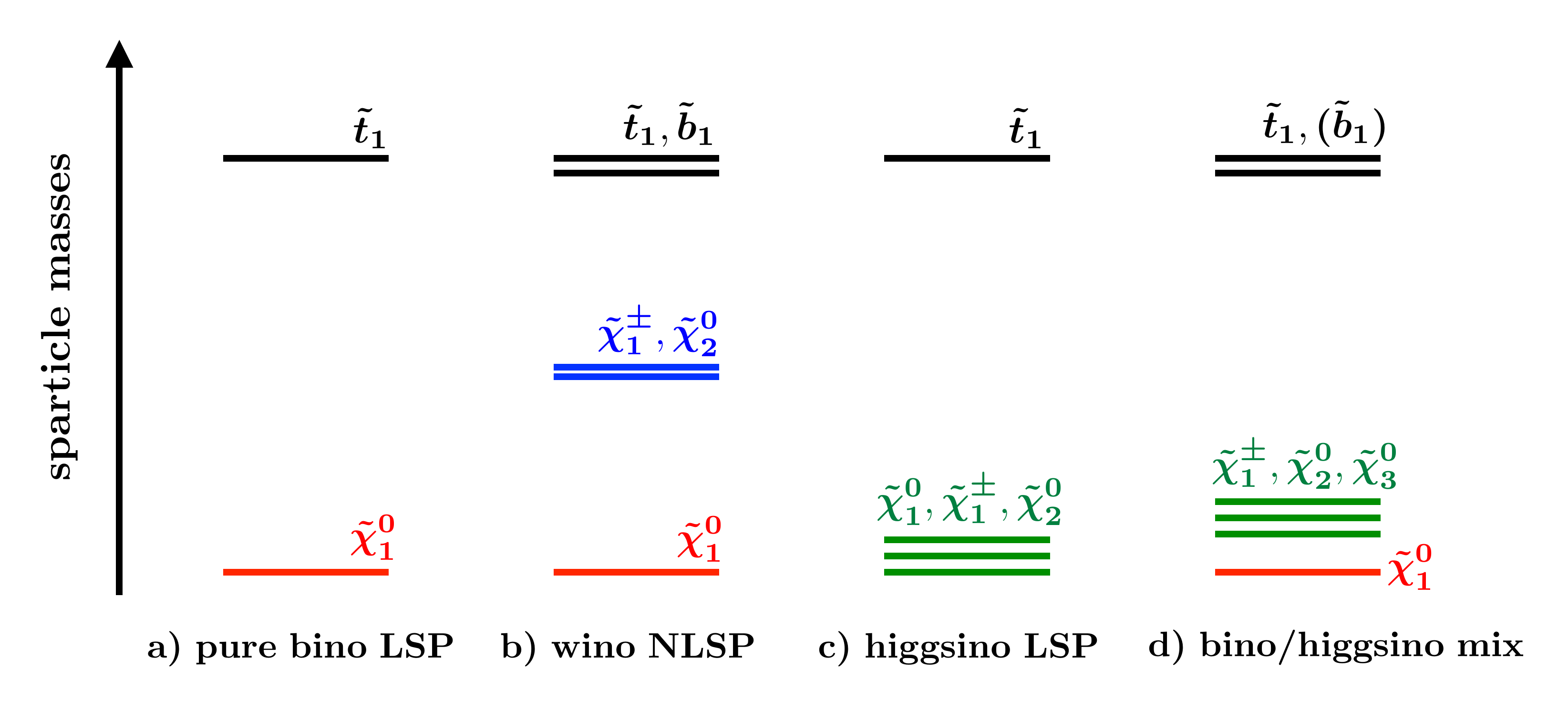}
\caption{Illustration of the sparticle mass spectrum for various LSP scenarios~\cite{ATL-CONF-2017-037}:
 a) Pure bino LSP, b) wino NLSP, c) higgsino LSP, and d) bino/higgsino mix. The \tone\ and \bone, shown as black lines, decay to various electroweakino states: the bino state (red lines), wino state (blue lines), or higgsino state (green lines), possibly with the subsequent decay into the LSP. The light sbottom (\bone) is considered only for pMSSM models with $m(\tleft)<$ $m(\tright)$.}
\label{fig:sparticle_mass_spectrum}
\end{center}
\end{figure}

\hspace{-0.60cm}
Four LSP scenarios are considered, with the collider signature dictated by the nature of the LSP: (a) pure bino LSP, (b) bino LSP with a wino next-to-lightest supersymmetric particle (NLSP), (c) higgsino LSP, and (d) mixed bino/higgsino LSP. The scenarios are detailed below with the corresponding sparticle mass spectra illustrated in Figure~\ref{fig:sparticle_mass_spectrum}. Complementary searches targeting scenarios where the LSP is a pure wino (yielding a disappearing track signature common in anomaly-mediated models of SUSY breaking) as well as other LSP hypotheses (such as gauge-mediated models) are not discussed further here.

\begin{enumerate}[label=(\alph*)]

\item Pure bino LSP model:

A simplified model is considered for the scenario where the only light sparticles are the stop (composed mainly of \tright) and the lightest neutralino. When the stop mass is greater than the sum of the top-quark and the LSP masses, the dominant decay channel is via \topLSP. If this decay is kinematically disallowed, the stop can undergo a three-body decay, \threeBody\, when the stop mass is above the sum of masses of the bottom-quark, $W$-boson, and $\ninoone$. Otherwise the decay proceeds via a four-body process, $\fourBody$, where $f$ and $f'$ are two distinct fermions, or via a flavour-changing neutral current (FCNC) process, such as the loop-suppressed \charmDecay. Given the very different final state, the FCNC decay is not considered further in the searches described below. The region of phase-space along the line of $m_{\tone} = m_{\ninoone} + m_t$ is especially challenging to target because of the similarity of the stop signature to the $\ttbar$ process, and is referred to in the following as the `diagonal region'.

\item Wino NLSP model:

A pMSSM model is designed such that a wino-like chargino (\chinoonepm) and neutralino (\ninotwo) are mass-degenerate, with the bino as the LSP. This scenario is motivated by models with gauge unification at the GUT scale such as the cMSSM or mSugra, where $M_2$ is assumed to be twice as large as $M_1$, leading to the \chinoonepm\ and \ninotwo\ having masses nearly twice as large as that of the bino-like LSP.

In this scenario, additional decay modes for the stop (composed mainly of \tleft) become relevant, such as the decay to a bottom-quark and the lightest chargino (\bChargino) or the decay to a top-quark and the second neutralino (\topNLSP). The $\chinoonepm$ and $\ninotwo$ subsequently decay to $\ninoone$ via emission of a (potentially off-shell) $W$-boson or $Z$/Higgs ($h$) boson, respectively. 

\item Higgsino LSP model:

`Natural' models of SUSY suggest low-mass stops and a higgsino-like LSP. In such scenarios, the typical mass splitting ($\Delta m$) between the LSP and \chinoonepm\ varies between a few hundred \MeV\ to several tens of \GeV\, depending mainly on the mass relation amongst the electroweakinos. For this analysis, a simplified model is designed for various $\Delta m(\chinoonepm,\ninoone)$ of up to 30 \GeV\ satisfying the mass relation:
\[
\Delta m(\chinoonepm,\ninoone) = 0.5 \times \Delta m(\ninotwo,\ninoone).
\]

The stop decays into either $b \chinoonepm$, $t \ninoone$, or $t \ninotwo$, followed by the \chinoonepm\ and \ninotwo\ decay through the emission of a highly off-shell $W/Z$ boson. Hence the signature is characterised by low-momentum objects from off-shell $W/Z$ bosons, and the analysis benefits from reconstructing low-momentum leptons (referred to as soft-leptons). The stop decay BR strongly depends on the \tright\ and \tleft\ composition of the stop. Stops composed mainly of \tright\ have a large branching fraction to \bChargino, whereas stops composed mainly of \tleft\ decay mostly into $t \ninoone$ or $t \ninotwo$. In these searches, both scenarios are considered separately.

\item Bino/higgsino mix model:

The `Well-tempered Neutralino' scenario seeks to provide a viable dark matter candidate while simultaneously addressing the problem of naturalness by targeting a LSP that is an admixture of bino and higgsino.  The mass spectrum of the electroweakinos (higgsinos and bino) is expected to be slightly compressed, with a typical mass splitting between the bino and higgsino states of $20$-$50$\,$\GeV$. A pMSSM signal model is designed such that low fine-tuning of the pMSSM parameters is satisfied and the annihilation rate of neutralinos is consistent with the observed dark matter relic density, $\Omega h^2$, where $\Omega$ is the density parameter and $h$ is Hubble constant ($0.10<\Omega h^2<0.12$).

\end{enumerate}

\subsection{Event selection}

Dedicated analyses are designed to achieve sensitivity to the broad range of scenarios mentioned above. Each of these analyses corresponds to a set of event selection criteria, referred to as a signal region (SR). All regions are required to have isolated lepton(s), jets, and large $\met$. In most cases, at least one $b$-tagged jet is also required. A set of preselection criteria is defined to monitor the modelling of the kinematic variables in simulated events. The preselection criteria are also used as the starting point for the SR optimisation. 
For the 1-lepton analysis, in order to reject multijet events, requirements are imposed on the transverse mass ($\mt$) and the azimuthal angles between the leading and sub-leading jets and $\met$. For the 2-lepton analysis, oppositely charged leptons are required with an invariant mass greater than 20 GeV in order to remove leptons originating from low mass resonances.

\subsubsection{Discriminating variables}

The backgrounds after preselection are dominated by the (semi-leptonic and dileptonic) \ttbar\ process. Additional discriminating variables help reduce the backgrounds further while retaining signals. The \mtTwo~\cite{Lester:1999tx} variables are widely used in stop searches, generalising transverse mass to signatures with two particles that are not directly detected. The \amtTwo\ variable targets dileptonic \ttbar\ events where one lepton is not reconstructed, while the \mtTwoTau variable targets \ttbar\ events where one of the two $W$-bosons decays via a hadronically decaying $\tau$ lepton. For the 1-lepton analysis, reconstructing the hadronic top-quark decay (top-tagging) can provide additional discrimination against dileptonic \ttbar\ events, which do not contain a hadronically decaying top-quark. In the diagonal region where $m_{\tone} \approx m_t + m_{\ninoone}$, the momentum transfer from the $\tone$ to the $\ninoone$ is small, and the stop signal has very similar kinematics to the \ttbar\ process. In order to achieve good signal-to-background separation, a boosted decision tree (BDT) technique is employed. Furthermore the recursive jigsaw reconstruction (RJR) technique~\cite{Jackson:2016mfb} is employed, which defines a new set of observables based on the assignment of physics objects in the event to the ``initial-state-radiation'' and ``sparticle'' systems. The RJR variables are used as input of the BDT analysis.

\subsection{Background estimation}

The main background processes after the signal selections include $\ttbar$, single-top $Wt$, $\ttbar+Z(\rightarrow \nu\nu)$, $W$+jets and diboson processes. Each of these SM processes are estimated by building dedicated control regions (CRs) enhanced in each of the processes, making the analysis more robust against potential mis-modelling effects in simulated events and reducing the uncertainties on the background estimates. The backgrounds are then simultaneously normalised in data for each SR with their associated CRs. The background modelling as predicted by the fits is tested in a series of validation regions (VRs). Systematic uncertainties due to theoretical and experimental effects are considered for all background processes and extrapolated to each SR.

\subsection{Results}

\begin{figure}[htbp]
        \centering
        \includegraphics[width=.50\textwidth]{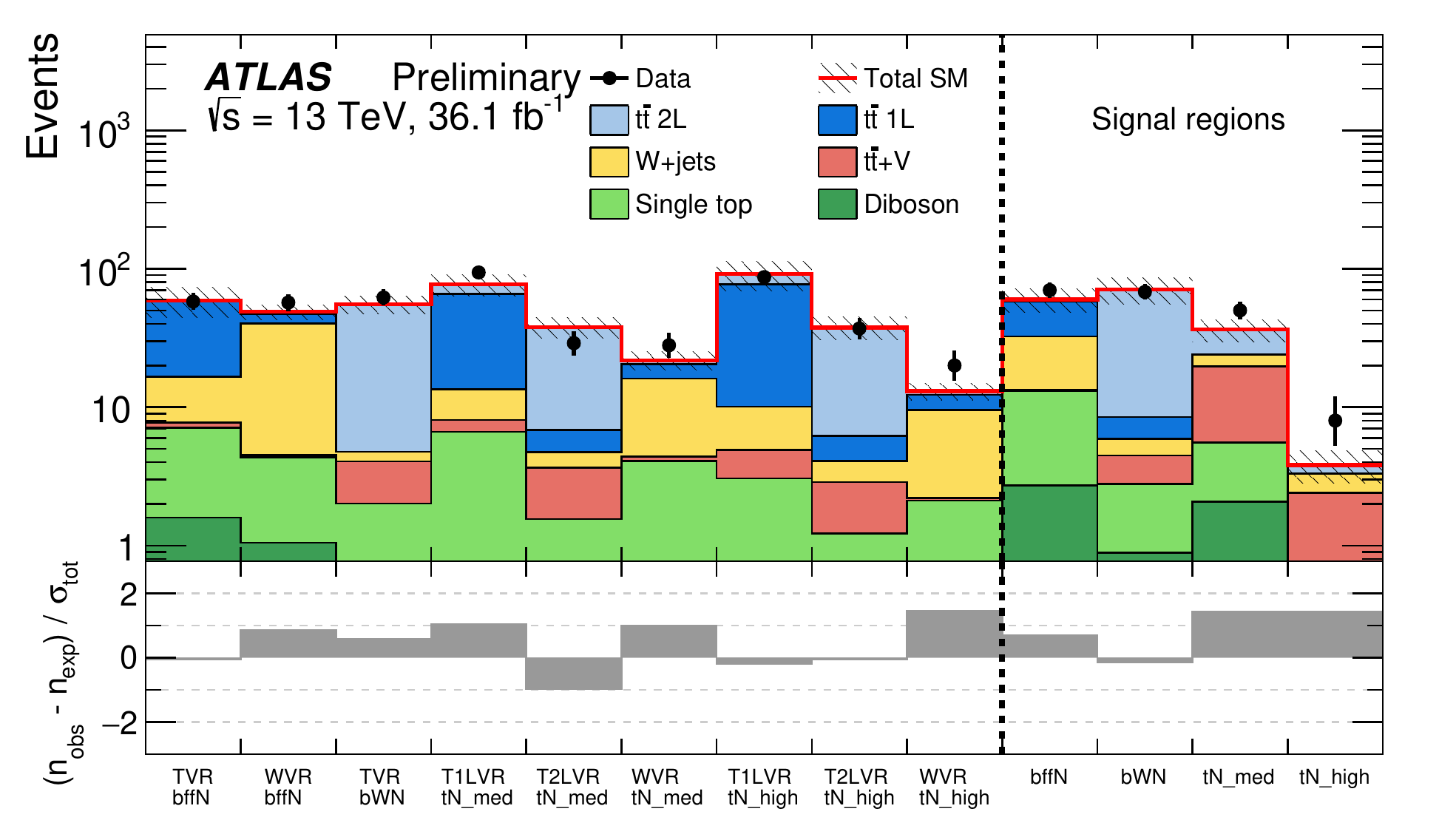} \\
        \caption{Comparison of the observed data ($n_\text{obs}$) with the predicted SM background ($n_\text{exp}$) in the signal regions designed targeting the bino LSP scenario~\cite{ATL-CONF-2017-037}. The background predictions are obtained using the background-only fit configuration, and the hashed area around the SM prediction includes all uncertainties. The bottom panels show the difference between data and the predicted SM background divided by the total uncertainty ($\sigma_\text{tot}$).
        }
        \label{fig:pulls-tN}
\end{figure}

After the determination of the SM background yield in the SR using the simultaneous fit (the so-called background-only fit that includes various CRs but not the SR), the number of observed data events in the SRs are compared to the predicted background yields. Figure~\ref{fig:pulls-tN} shows comparisons between the observed data and the SM background prediction in the SRs from the stop 1-lepton analysis.

\vspace{0.20cm}

\begin{figure}[htbp]
        \centering
        \includegraphics[width=.35\textwidth]{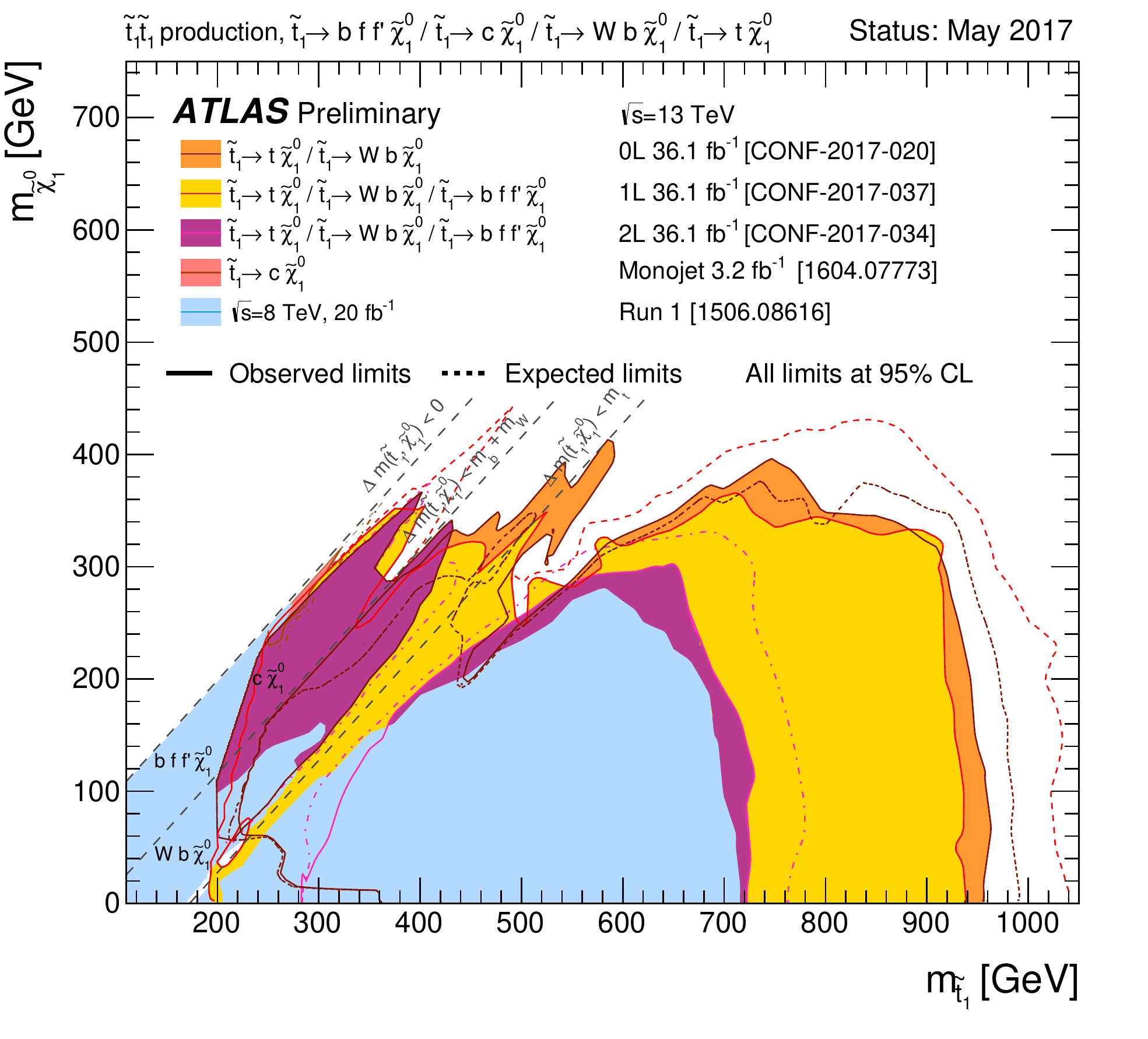}%
        \includegraphics[width=.45\textwidth]{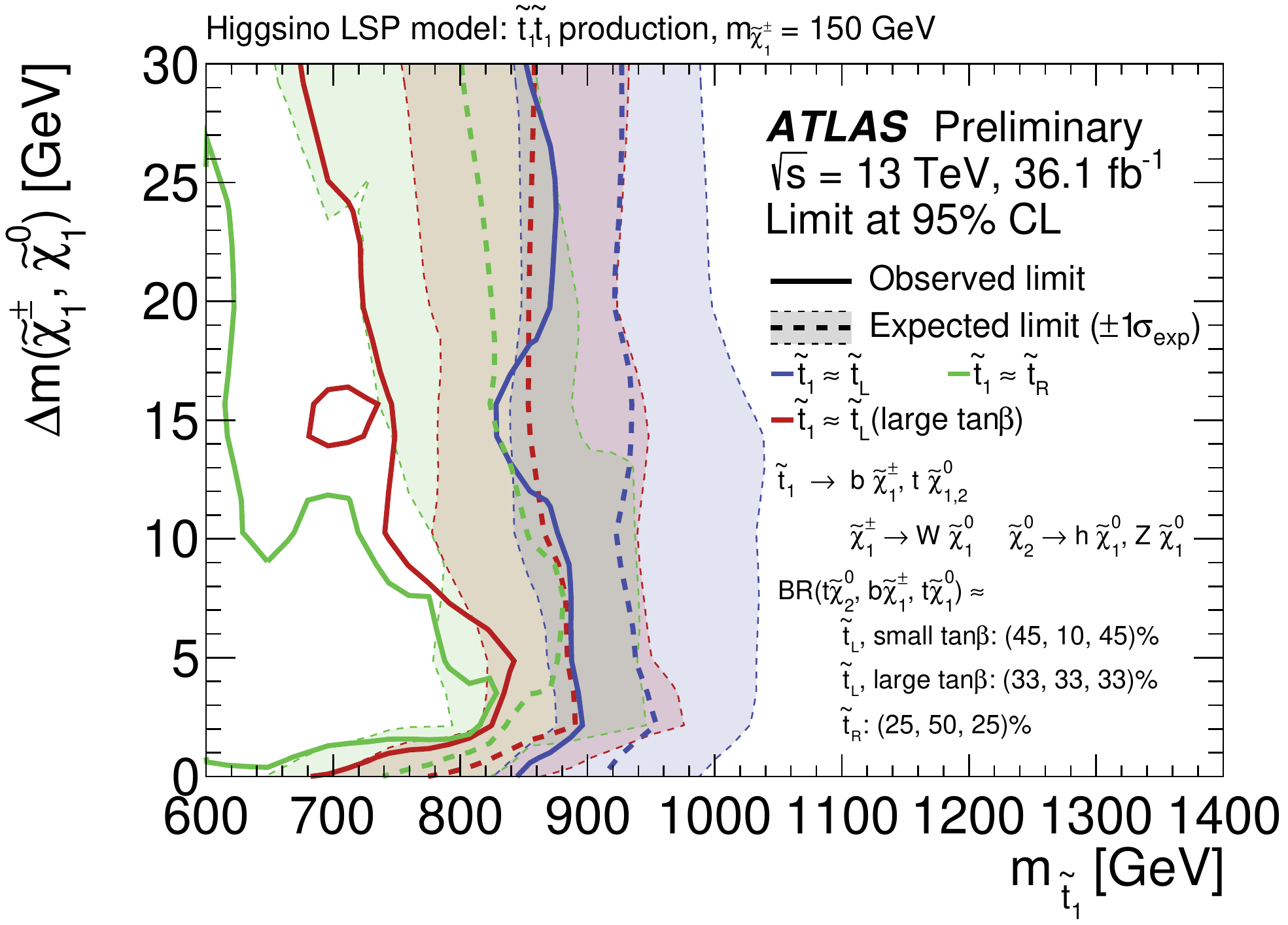}
        \caption{Expected (dash) and observed (solid) 95\% excluded regions: (left) the exclusion limits in the plane of $m_{\tone}$ versus $m_{\ninoone}$ for direct stop pair production assuming either $\topLSP$, $\threeBody$, or $\fourBody$ decay with a branching ratio of 100\% in the pure bino LSP model. In addition to the stop 1-lepton result, results from the stop 0-lepton and 2-lepton analysis are overlaid~\cite{ATL-CONF-2017-034,ATL-CONF-2017-020,ATL-SUSYPublic}.
 (right) the exclusion limits in the plane of $m_{\tone}$ versus $\Delta m$ ($\chinoonepm$, $\ninoone$) for direct stop pair production decaying into the higgsino LSP~\cite{ATL-CONF-2017-037}. 
        }
        \label{fig:stop-summary}
\end{figure}

\hspace{-0.60cm}
As no significant excess is observed, exclusion limits are set based on profile-likelihood fits for stop pair production models. Figures~\ref{fig:stop-summary} shows the expected and observed exclusion contours as a function of stop and neutralino mass for the pure bino LSP scenario and the higgsino LSP scenario. For the pure bino LSP scenario, the exclusion limits are obtained under the hypothesis of mostly right-handed stops.
For the higgsino scenarios, various stop polarisation scenarios that significantly affect the stop decay branching ratio are considered and overlaid. The stop masses of up to 940 GeV are excluded for the pure bino LSP model with massless neutralino and up to 900 GeV in the higgsino LSP model.

%

\section{$R$-parity violating scenarios}

General MSSM models may allow for violation of baryon number (B) and lepton number (L), leading to rapid proton decay. A common theoretical approach to evade the strong constraints from the non-observation of these processes is to introduce R-parity conservation. However a number of theoretical models predict R-parity violating (RPV) process, while protecting the proton from decaying. ATLAS searches define several benchmark RPV SUSY models, including two RPV stop searches presented below. Schematic diagrams are shown in Figure~\ref{fig:RPVstop}. 
\begin{figure}[htbp]
\begin{center}
	\vspace{-1.0cm}
	\includegraphics[width=0.50\textwidth]{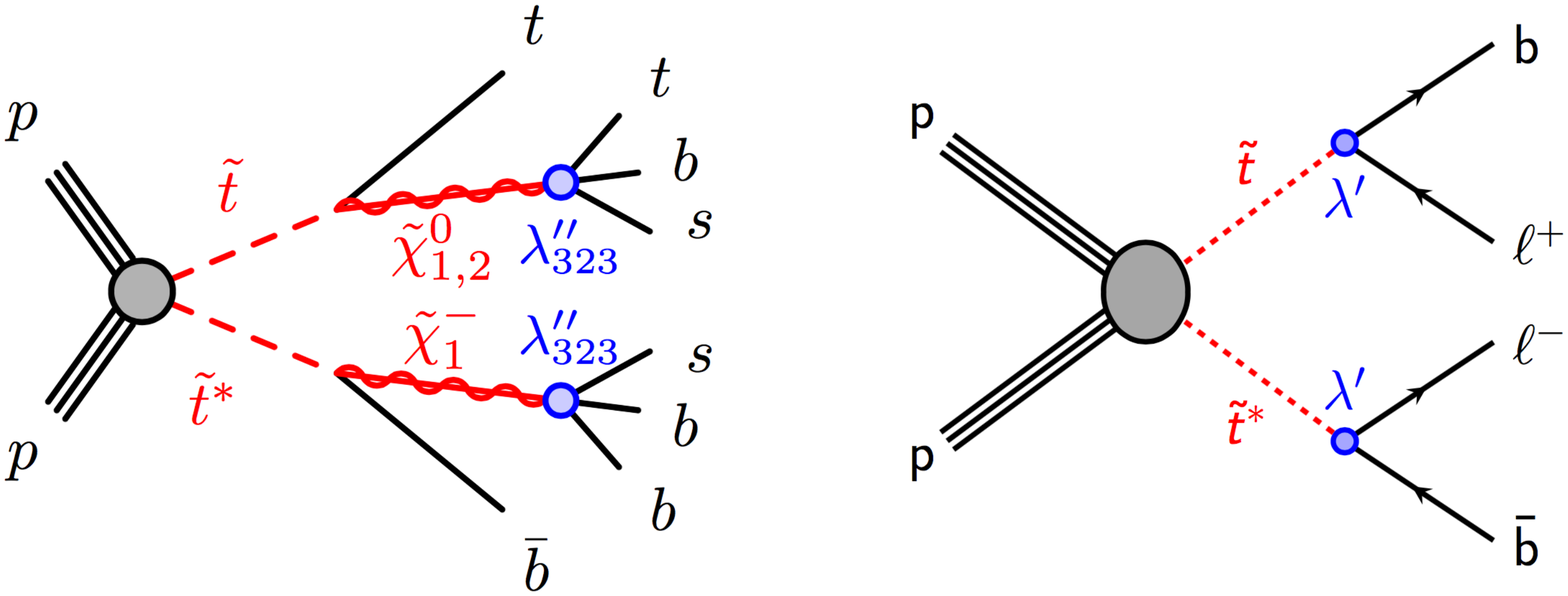}
	\vspace{-1.0cm}
	\caption{Illustration of the diagrams for direct stop pair production decaying via RPV couplings. The left diagram shows a benchmark model considered for the RPV 1-lepton analysis and the right shows a benchmark model considered for the stop B-L analysis~\cite{RPV1L,ATL-CONF-2017-036}.
	}
	\label{fig:RPVstop}
\end{center}
\end{figure}

\vspace{-0.40cm}
\hspace{-0.60cm}
\paragraph{RPV 1-lepton model~\cite{RPV1L}:} This scenario assumes the pair production of right-handed stops, which decay to bino or higgsino LSPs. The LSP (lightest neutralino) decay proceeds through the non-zero RPV couplings (e.g. $\lambda^{''}_{323}$) allowing for the prompt decay of the LSPs. The signature is characterised by up to 12 jets including $b$-tagged jets and one isolated lepton from top-quark decays. The dominant background in the SR is the \ttbar\ process. A dedicated data-driven parametrized template technique is established to estimate the background at high jet multiplicity.
\vspace{-0.40cm}

\hspace{-0.60cm}
\paragraph{stop B-L model~\cite{ATL-CONF-2017-036}:} This model introduces an additional local symmetry $U(1)_{B-L}$ to the SM with right-handed neutrino supermultiplets. The B-L symmetry is then spontaneously broken by a right-handed sneutrino. This minimal B-L model violates L but not B. In this model the stops are pair produced and the stop decay proceeds via RPV couplings ($\lambda^{'}$), which are highly suppressed as they are related to the neutrino masses, carrying color and electric charge. The analysis targets a final state of two bottom quarks and 2-leptons and searches for a bump in the invariant masses of the jet-lepton pair. Therefore the lepton-jet invariant mass and contransverse mass ($m_{CT}$) are used as discriminating variables. The backgrounds in the SR are dominated by the single-top $Wt$ and \ttbar\ processes. Dedicated single-top and \ttbar\ CRs are defined to determine the normalisation of the backgrounds in the SR.

\subsection{Results}

\begin{figure}[htbp]
        \centering
        \includegraphics[width=.35\textwidth]{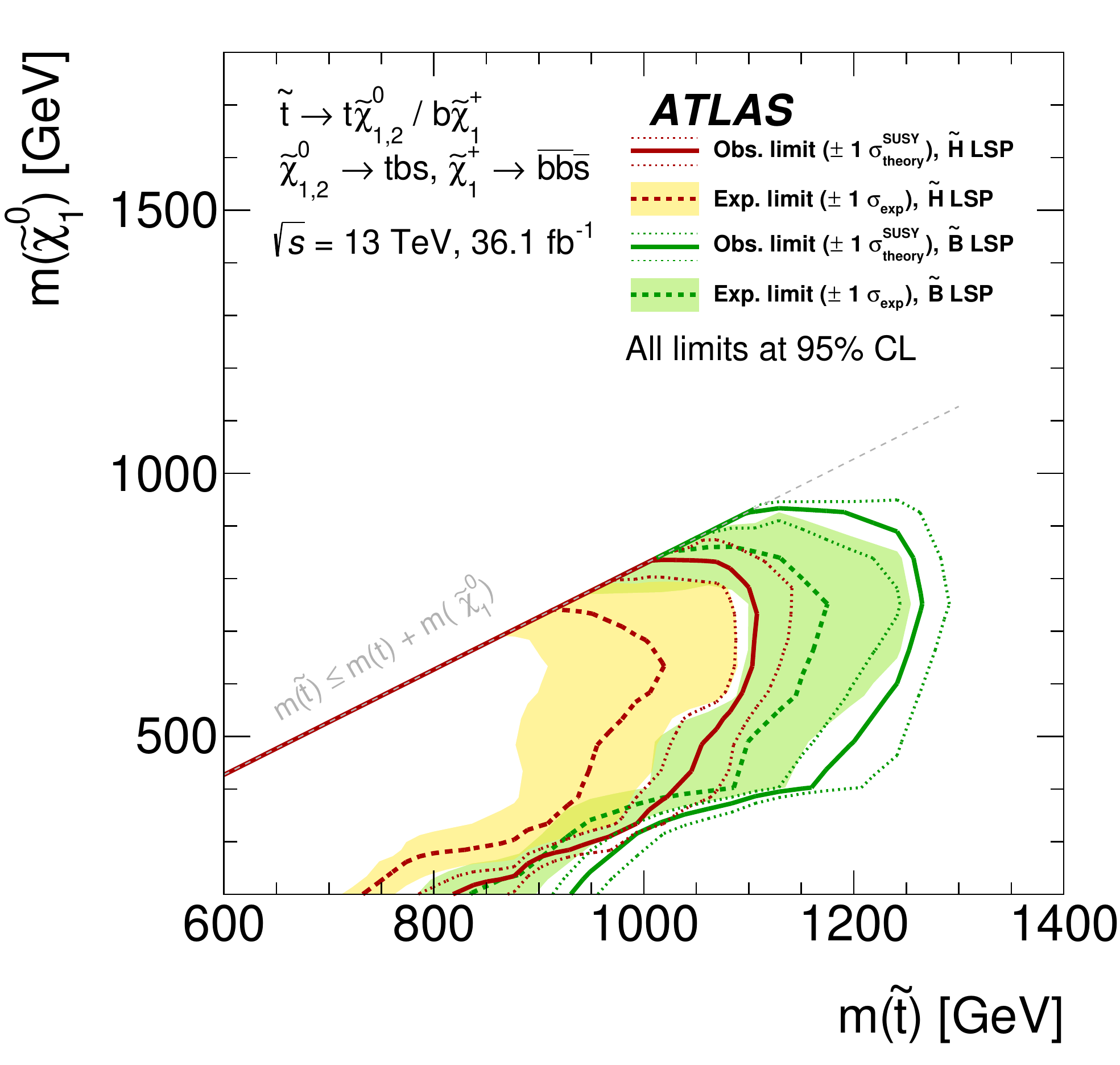}
        \includegraphics[width=.45\textwidth]{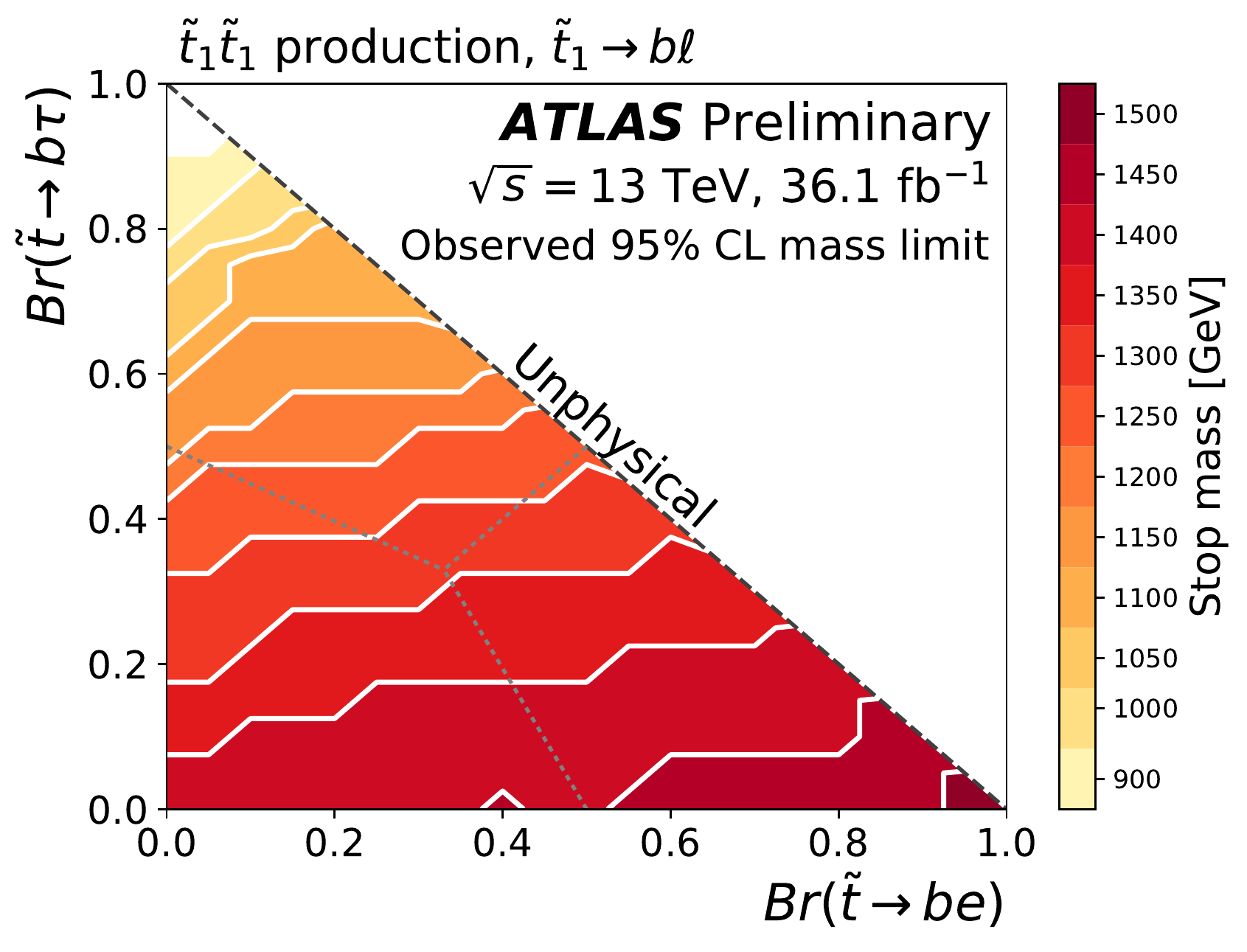}%
        \caption{Expected (dash) and observed (solid) 95\% excluded regions: (left) in the plane of $m_{\tone}$ versus $m_{\ninoone}$ for direct stop pair production in the bino LSP and higgsino LSP models, obtained by the RPV 1-lepton analysis~\cite{RPV1L}, and (right) in the plane of electron channel BR versus muon channel BR for direct stop pair production in the stop B-L model~\cite{ATL-CONF-2017-036}.
        }
        \label{fig:RPVstop-summary}
\end{figure}

As no significant excess is observed, exclusion limits are set for stop pair production models decaying via RPV couplings as shown in Figure~\ref{fig:RPVstop-summary}. For the $\lambda^{''}_{323}$ model, a stop mass up to 1.2 TeV is excluded depending on the type of LSP. For the stop B-L model, strong limits are obtained for the various BR scenarios. A stop mass up to 1.5 TeV is excluded depending on the stop branching fraction.

\section{Conclusions}

Third-generation squarks have been extensively searched for in the ATLAS experiment and analyses in the leptonic final state are presented. No significant excesses of observed data over predicted background processes have been observed, and exclusion limits have been set on various stop models. For the R-parity conserved models, a stop mass up to 940 GeV has been excluded in various LSP scenarios. For the R-parity violating scenario, a stop mass up to 1.5 TeV has been excluded depending on the RPV couplings considered. 



\end{document}

%% file: econfmacros.tex
\def\beq{\begin{equation}}
\def\eeq#1{\label{#1}\end{equation}}
\def\eeqn{\end{equation}}

\def\beqa{\begin{eqnarray}}
\def\eeqa#1{\label{#1}\end{eqnarray}}
\def\eeqan{\end{eqnarray}}

\def\Dslash{\not{\hbox{\kern-4pt $D$}}}
\def\dslash{\not{\hbox{\kern-2pt $\del$}}}

\def\Pl{{\mbox{\scriptsize Pl}}}

\def\mt{m_t}

\def\msb{{\bar{\ssstyle M \kern -1pt S}}}

